\documentstyle [12pt]   {article}
 \evensidemargin\oddsidemargin
\begin{document}
\count0 = 1
 \title{ Information  Constraints in Quantum\\
 Measurements  and State Collapse   }
\author{ S.N.Mayburov \\
Lebedev Inst. of Physics\\
Leninsky Prospect 53, Moscow, Russia, 117924\\
E-mail :\quad   mayburov@sci.lebedev.ru}
\date {}
\maketitle
 \vspace {10mm}
 \vspace{10mm}

\begin{abstract}

 Quantum-mechanical constraints on 
 information transfer in measuring  systems and their influence on
 measurement results  studied. As the  example,  measurement of
binary observable $S_z$ of   object $\cal S$ 
  by measuring apparatus $\cal A$
  considered.  
  It's shown that during the measurement these constraints  obstacle 
    the acquisition of information by $\cal A$,   characterizing purity of $\cal S$  ensemble.
 Due to  it, $\cal A$ can't discriminate
 pure and mixed $\cal S$ ensembles with the same $S_z$ expectation value.
In algebraic measurement  ansatz by Emch  such information loss results in  stochastic $S_z$ measurement outcomes for pure $\cal S$ ensemble,    their probabilities obey to Born postulate, i.e. it corresponds to  quantum  state collapse. 
Account  of $\cal A$ state decoherence  doesn't  change  obtained results principally.
\end{abstract}


\section  {Introduction}

 Quantum mechanics (QM) after more than hundred years of its
development  achieved  tremendous success in the description
of nature, there are serious reasons to assume that QM is universally valid [1-3]. However, its foundations are still disputed extensively
 and  contain some unsettled aspects up to now.
The problem of quantum measurement (state collapse), i.e.  randomness of state superposition measurement outcomes is one of such topics. In fact, the main question behind it is, whether it's possible to derive Born measurement postulate (Luder's rule) from other QM premises without accepting it as axiom
or only some principally different physical theory can explain it. The list of books and reviews devoted to measurement problem is quite extensive ([1-4] and refs. therein). In the last years significant attention attracts its study in the framework of  microscopic measurement models, such models permitted to obtain  important results for   some other aspects of quantum measurements [1,2]. In  its typical experimental scheme the  measuring system $\cal T$  includes  studied object  $\cal S$, which interacts with detector $D$, as the result of this interaction, its final state correlates with $\cal S$ state and can amplify some its parameters \cite {B1,Es2,M6}. In its turn $D$ interacts with information gaining and utilizing  system (IGUS) $\cal I$, which stores and processes information about  $D$ state, consequently, its final state also correlates with initial $\cal S$ state.
 In such models $D$ and
  $\cal I$  treated as  quantum objects and
   described  by  density matrix $\rho(t)$,
 the  measurement process described by standard (Dirac) QM formalism  \cite {J3}. It disputed still whether arbitrary  $D$ and $\cal I$ can be considered as quantum objects, or some kind of
Heisenberg cut should be imposed inside measuring system dividing it into quantum and classical (macroscopic) parts \cite {B1,Es2,M6}. However,  no watertight arguments or experimental results in favor of its existence were obtained up to now, hence such systematically quantum approach  to measurement process  looks consistent. It permits also to account $D$ and $\cal I$ decoherence by their environment [1-2].

In this paper the quantum measurements will be considered in the framework of quantum information theory. Really, since any  measurement is transfer of information on  measured $\cal S$ state parameters to $\cal I$, then the measuring system can be  treated also as an information channel \cite {M8,M9}.  It was shown earlier that in comparison with classical channels  such quantum channels possess essentially smaller capacity \cite {Barn,HHH}.  Therefore, it's possible, in principle,  that such capacity limitations would result in  the information losses during $\cal S$ signal  transfer, and thus  can influence $\cal S$ measurement outcome for $\cal I$.
However, until now their influence on
quantum measurement characteristics weren't analyzed thoroughly.
 In our previous papers it was shown for  simple models that during $\cal S$ observable measurements
such limitations result in  the loss of
information, characterizing   $\cal S$ ensemble purity rate, so
it becomes impossible for $\cal I$ to discriminate  pure and mixed
ensembles of incoming $\cal S$ states,  at least at the level of  observable expectation values \cite {M8,M9}. Here this approach will be extended on the description of measurement outcomes  for more realistic measurement models and elaborated mathematical formalism.  


Standard simplification exploited in   microscopic models is  to replace $D$ and $\cal I$ by single object - – the measuring apparatus $\cal A$, which stores and processes the  information on $\cal S$ measurement outcomes \cite {B1,M6}.  
In this measurement scheme, due to the choice of $\cal  S$, $\cal A$ interaction Hamiltonian, for $\cal S$ eigenstate superposition the corresponding $\cal T$ evolution  results in $\cal S$, $\cal A$ final state entanglement. It means that  $\cal A$ state can't be factorized from $\cal S$ state and consequently from $\cal T$ state.  Due to it,
$\cal A$ can be formally treated as $\cal T$ subsystem, in other words, $\cal A$ measures the state of system of which $\cal A$ is its part. It was argued earlier  that such situation can be treated formally as  particular case of  self-reference problem, well-known in computability and information theories (\cite {M6,S7} and refs. therein). General properties of   self-referential measurement formalism  for classical and quantum systems were studied in \cite {B4,B5}. 
It was shown that  for given $\cal S$  state measurement the resulting  $\cal A$ state defined by corresponding $\cal T$ state map to $\cal A$  degrees of freedom. However,  standard QM formalism  doesn't permit to calculate this map unambiguously,  proposed 'ad hoc' variant leads to  controversial  results  \cite {B5}.  

 It will be shown here that  algebraic QM formalism  permits to derive consistent $\cal T$   state mapping to $\cal A$.
 Operator algebras universally appreciated now as well-founded generalization of standard QM formalism [12-14]. Its application to quantum measurements started from seminal paper by Hepp \cite {He}. Standard algebraic approach to measurements assumes  that $\cal A$ or its environment can be treated as (nearly)
infinite  systems (\cite {Lan,B6} and refs. therein).  However, this approach was shown to possess serious loopholes \cite {B1,M6}. Here auxilary  ansatz of partial states by Emch will be applied to quantum  measurements
 \cite {Em4,Em5}. It provides unambiguous  measurement description both for finite and infinite systems. In its framework, for  $\cal S$ eigenstate superposition the final $\cal A$ states correspond to stochastic  ensemble of 
individual states, their  probabilities correspond to Born postulate.
  Effects of measuring system decoherence by its environment
 can be also accounted, 
it  will be studied  here for Zurek model in  sect. 6 $\cite {Z10}$.


\section         {Microscopic Measurement  Model}

Here  some  aspects of
QM measurement theory, important for our study,   reviewed.
In the ideal scheme of state preparation the individual quantum states can be prepared event by event by experimenter. In particular, it assumes that experimenter by his choice can prepare the superposition of  orthogonal pure states or either of those states with some probabilities. Described ensembles are  just particular cases of all  possible  quantum ensembles, yet at first only these options, often applied for measurement studies, will be considered \cite {B1,B4}.
In QM formalism the individual system  states are pure, i.e. extremal, normalized, positive functionals on system observable set or algebra. They are described as Dirac vectors $|\psi\rangle$
or, equivalently, projectors $P=|\psi\rangle\langle\psi|$  on  system Hilbert space $\cal H$.
Statistical  states, describing quantum ensembles, are $\sigma$-weakly continuous, i.e. normal, normalized, positive functionals, they correspond to  density matrixes $\rho$ on $\cal H$ \cite {J3,Em4}.  Individual states constitute the shell of statistical state set.
 When the
 statistical ensemble preparation performed according to described procedures, their
decomposition into  individual state set can admit more detailed
description in form of gemenge \cite {B1}. For statistical mixture of binary states  considered here, it described as
 ensemble of  pure states with $P_{1,2}$ projectors prepared by experimenter with
probabilities $w_{1,2}$, such  that $P_1+P_2=I$ and $ w_1+w_2=1$.
This  ensemble is described  by gemenge array
$W=\{w_i; P_i\};\, i=1,2$,  where $P_{1,2}=|\psi_{1,2}\rangle\langle \psi_{1,2}|$,
 its  density matrix  $\rho_g= w_1P_1+w_2 P_2$.




%


 We'll exploit here Coleman-Hepp model for $\cal S$, $\cal A$  evolution  description, it provides reasonable imitation of macroscopic apparatus $\cal A$ functioning \cite {He,Bell}.   In its framework,  $\cal S$ is  the particle $\cal P$ with the spin-$\frac{1}{2}$,
 its up, down states
 denoted $|s^0_{1,2}\rangle$,
  initial $\cal P$ spin state is:
\begin {equation}
 |\psi_s\rangle =e_1|s^0_1\rangle+e_2|s^0_2\rangle
  \label {AA99}
\end {equation}
for arbitrary complex $e_{1,2}$ with $|e_1|^2+|e_2|^2=1$.
For the comparison,  measurement of
  $|s^0_{1,2}\rangle$   statistical ensemble  will be studied, it described by  gemenge
$W^s=\{w_i, |s^0_i\rangle \langle s^0_i|\}   ; i=1,2$, here ${w}_i$ $=|e_i|^2$.

 Apparatus $\cal A$ is linear array (chain) of $N$ spin-$\frac{1}{2}$ particles
  $\cal C$$^i;$
   $i=1,...,N$ fixed    at positions $x_i=1,...,N$, $\cal P$ and $\cal A$ constitute the measuring system $\cal T$. 
  $\cal P$, $\cal A$ combined wave function is $\psi(t,x,s^0,{.}{.}{.},s^i, {.}{.}{.})\,;i=1,...,N$,
where  $s^0, s^i$ are $\cal P$, $\cal C$$^i$ spin projections on $Z$ axe,     
  $x$ is $\cal P$ coordinate.
 $\sigma^0_{x,y,z}, \sigma_{x,y,z}^i$ are $\cal P$, $\cal C$$^i$ Pauli matrixes \cite {J3}, corresponding $\cal P$, $\cal C$$^i$ spin  operators denoted $S^0_{x,y,z}, S_{x,y,z}^i$; $\, i=1, ...,N$.
Initially, at $t=0$ 
 all $\cal C$$^i$  are in up spin state $|s^i_1\rangle$; their down states denoted  $|s^i_2\rangle$, corresponding polarized $\cal A$ 'pointer' states
 exploited in this model 
\begin {equation}
|A_{1,2}\rangle= \prod_{i=1}^{N} \otimes |s^i_{1,2}\rangle \label {Z0}
\end {equation}
  $\cal P$, $\cal A$ spin Hilbert spaces
 denoted as  $\cal H$$_s$, $\cal H$$_{ A}$ correspondingly.  In this set-up at $t=0$ $\cal P$ wave packet starts to move in $X$ direction along $\cal A$ spin chain and reaches its end at $t=t_f$.
 Then, for Coleman-Hepp $\cal P$, $\cal A$ Hamiltonian $H_{P,A}$ the solution of Schroedinger equation  indicates that
 after $\cal P$ passage along $\cal A$ spin chain at $t >> t_f$, $\cal P$, $\cal A$  spin state becomes entangled and factorized from its $x$-coordinate part \cite {He}
\begin {equation}
    \psi(t,x,s^0,s^{1},...,s^{i},...,s^{N}) =\chi(x,t)[e_1|s^0_1\rangle|A_1\rangle+
(-i)^N e_2|s^0_2\rangle|A_2\rangle]   \label {Z2}
\end {equation}
 Obtained final $\cal T$ state $\psi$ has the interpretation that when $\cal P$ spin is up,
 nothing happens to $\cal A$  initial spin orientation,
  but when it is down then all $\cal A$ spins  flipped from up to down,
   so that for $\cal P$ spin state superposition it results in $\cal P$, $\cal A$ state entanglement.
For incoming $\cal P$ gemenge the resulting ensemble will be
described by joint $\cal P$, $\cal A$ gemenge $W^{\cal T}$$=\{w_i,
|s^0_i\rangle \langle s^0_i||A_i\rangle \langle A_i|\}$;  i=1,2. In such model 
only
 $S^0_z$ eigenstates can be measured by $\cal A$ unambiguously resulting in $|A_{1,2}\rangle $ final $\cal A $ states, they 
 constitute in $\cal H$$_A$ the pointer
basis  $|A_{1,2}\rangle$. Corresponding $\cal A$ pointer observable 
 \begin {equation}
S^A_z=\sum_{i=1}^N S^i_z =\frac{1}{2}\sum_{i=1}^N \sigma^i_z \label {Z55}
  \end {equation}
  its final expectation value is equal to $N\bar{S}^0_z$  for arbitrary $\cal P$ state,   conjugated $\cal A$ observables $S^A_{x,y}$ defined analogously.  

It's worth
to notice beforehand that even if  $\cal A$  supposed to be the human
brain, in our approach the observer's consciousness doesn't play
any role and will not be referred to at all. 
 However, for  illustrative purposes  some terms characteristic for
 conscious analysis of acquired $\cal A$  signals will be used  here.

\section {Information transfer in quantum measurements}



  As was argued above, 
 the measuring system $\cal T$ can be considered as the information  channel connecting measured object $\cal S$ and IGUS $\cal I$ \cite {M6}. It was demonstrated  earlier that  in comparison with classical channels, such quantum  channels possess essentially smaller capacity, hence, in principle, possible $\cal S$ information losses   can influence measurement outcome for $\cal I$
\cite {Barn,HHH}.
In our previous papers it was shown
for  simple measurement models that
such restrictions result in  the loss of
information, characterizing the  $\cal S$ ensemble purity rate, so
it becomes impossible for $\cal I$ to discriminate the pure and mixed
ensembles of incoming $\cal S$ states,  at least at the level of  observable expectation values \cite {M8,M9}. Here this approach will be applied to more  realistic measurement models. 

 For Coleman-Hepp model $D$ and $\cal I$ replaced by measurement
 apparatus $\cal A$. For  detailed study of measurement process
let's define first  the class of  interference observables $\{B \} $ which expectation values for spin-$\frac {1}{2}$ particle $\cal P$ can discriminate  the pure and mixed $\cal P$ ensembles with the same ${S}^0_z$ expectation value.
For that purpose rewrite $\psi_s$ of eq. (\ref {AA99}) as
\begin {equation}
|\psi_s\rangle=|e_1||s^0_1\rangle+ \exp(i\gamma)|e_2| |s^0_2\rangle \label {Z22}
\end {equation}
so that $\gamma$ is $e_{1,2}$ quantum phase difference.   In general such 
observables correspond to self-adjoint operators 
\begin {equation}
B(b_{1,2})=b_1|s^0_1 \rangle \langle s^0_2|+  b_2|s^0_2\rangle \langle s^0_1|   
\label {Z23}
\end {equation} 
with arbitrary  $b_1=b_2^*$, their norm can be chosen as $|b_{1,2}|=\frac{1}{2}$ \cite{Es2,M8}.  Then, for  arbitrary $\cal P$ mixed ensemble described by gemenge $W^s$  of sect. 2, it follows that
for arbitrary $b_{1,2}$ values $\bar B=0$. For pure ensemble $\bar{B}\ne 0$ for all  $b_{1,2}$ values, except
the case when 
$$
\frac{b_1}{b_2}=exp[i(\gamma+\frac{\pi}{2}+n\pi)] 
$$
with $n=0,\pm1,\pm2,...$. Interference term $\bar B$ acquires maximal value $|e_1||e_2|$ for
$\frac{b_1}{b_2}=exp[i(\gamma+2n\pi)]$. Hence by measuring $B$ observable with suitable $b_{1,2 }$ parameters  one can define purity rate of $\cal P$ ensemble \cite {B1}. It's notable, however, that arbitrary $B(b_{1,2})$ is conjugated to $S_z^0$, i.e. they obey to Heisenberg commutation relations which can induce  uncertainty relations for them during $\cal P$ state measurement.  

For $\cal P$, $\cal A$ entangled system its purity rate defined  analogously. 
Rewrite $\cal P$, $\cal A$ final  spin state of eq. (\ref {Z2}) as
\begin {equation}
|\psi\rangle=|e_1||s^0_1\rangle|A_1\rangle+ \exp(i \epsilon)|e_2| |s^0_2\rangle|A_2\rangle \label {Z24}
\end {equation} 
with $|A_{1,2}\rangle$ of eq. (\ref{Z0}), $\epsilon$ is quantum phase difference. Consider
first the case $N=1$,
then such observable class $\{G_1^s\}$ described as 
\begin {equation}
G_1^s(c_{1,2})= c_1| s^0_1\rangle \langle
s^0_2||s^1_1\rangle\langle s^1_2|+
c_2| s^0_2\rangle \langle
s^0_1||s^1_2\rangle \langle s^1_1|
\label {Z88}
\end {equation}
with arbitrary $c_2=c_1^*$,  their norm is chosen to be $|c_{1,2}|=\frac{1}{2}$. 
For arbitrary $N$ interference observables defined analogously:
\begin {equation}
G_N^s(c_{1,2})= c_1| s^0_1\rangle \langle
s^0_2||A_1\rangle\langle A_2|+
c_2| s^0_2\rangle \langle
s^0_1||A_2\rangle \langle A_1|
\label {Z91}
\end {equation}
For pure $\cal P$ spin states $\bar {G}_{N}^s$$\neq0$ except for
$$
\frac{c_2}{c_1}=exp[i(\epsilon+\frac{\pi}{2}+n\pi)]
$$ 
For initial  $|s^0_{1,2}\rangle$ gemenge $W^s$ the final $\cal P$, $\cal A$ ensemble will be mixed and for arbitrary $c_{1,2}$ it results in $\bar {G}_N^s=0$. Interference term $\bar {G}_N^s$ acquires maximal value $|e_1||e_2|$ for
$\frac{c_1}{c_2}=exp[i(\epsilon+2n\pi)]$. Hence, from the measurement of $G_N^s(c_{1,2})$ expectation value by  external apparatus $\cal A'$  one can define purity rate of joint  $\cal P$, $\cal A$ state
\cite {Lah}.
Meanwhile, it's easy to check that  no $\cal A$ observable expectation value  differs for  pure and mixed $\cal P$ ensembles with the same $\bar{S}^0_z$ value. For $N=1$ all $\cal A$ observables conjugated to $S^A_z$ can be written as $K^A=d_1 S^A_x + d_2 S^A_y$ for arbitrary real $d_{1,2}$.
Yet it follows that both for arbitrary pure and mixed $\cal P$ ensembles $\bar{K}^A=0$. Analogously to it, for arbitrary $N \geq 2$ only  $S^A_z$ can have nonzero expectation value for pure and mixed $\cal P$ ensembles, for all other $\cal A$ observables, conjugated to $S^A_z$  they are equal to  zero independently of incoming $\cal P$ ensemble.
It means that in $\cal P$ spin measurement process the information on $\cal P$ purity rate
doesn't transferred to $\cal A$, yet it does not disappear but transferred to $\cal T$ system as the whole (unbroken wholeness \cite {Es2}).  Hence due to this reason in such measurement $\cal A$ itself can't  discriminate pure and mixed $\cal P$ ensembles. Moreover, any measurement performed by some external observer on $\cal A$ also will not reveal the difference between mixed/pure $\cal P$ ensembles.
 In this analysis the effects of  $\cal A$ decoherence by its environment omitted, supposedly, they would make this conclusion even more stronger \cite {Z10}, because they would make the information about initial $\cal P$ state carried by $\cal A$ environment elements also (see below).  Obtained results demonstrate that at the level of statistical states  the apparatus $\cal A$ can't discriminate pure and mixed $\cal P$ ensembles.
 It evidences  that, in principle, the state collapse for individual states  can appear, yet  direct proof can be obtained only in algebraic QM formalism considered below.

\section {Self-reference in quantum measurements}

It admitted generally that complete QM formalism should describe also the measurement process, 
treating the measuring system as proper quantum system,  in that case QM universality would imply its semantical completeness \cite {M6,B4}. Its possible reliability  will be studied here for some  microscopic measurement models.  In these models the measuring system  $\cal T$ includes measured object $\cal S$ and measuring apparatus $\cal A$, which interacts with $S$  and stores the information on $\cal S$ state features.  For $\cal S$ eigenstate superposition the corresponding $\cal T$ evolution  results in $\cal S$, $\cal A$ final state entanglement, for Coleman-Hepp model it  described by eq. (\ref{Z2}).
As the result, final $\cal A$ state can't be factorized from $\cal S$ state and consequently from $\cal T$ state.  Due to it,
$\cal A$ can be formally treated as $\cal T$ subsystem, i.e. $\cal A$ measures the state of system of which $\cal A$ is its part.

It was suggested for long that such situation should be treated as particular case of self-reference problems well-known in mathematical foundations, computability and information theory \cite {S7,B4}. Its notorious example is Goedel theorem on formal system incompleteness,  its implications for self-referential measurements   analyzed in   \cite {S7,B4}.
 General properties of self-reference formalism  for such 'measurement from inside' for classical and quantum systems were studied by Breuer \cite {B4,B5}. 
It was shown that for realistic measuring system $\cal T$ the particular $\cal S$ measurement described by specific  mapping $M_S$ of arbitrary $\cal S$ state $\Gamma _S$ to  final $\cal T$  state  $\Gamma_T$, in the same time,  it induces the corresponding   $\Gamma_T$ mapping  $M_A$ of $\cal T$ state to apparatus $\cal A$ state $R_A$. Such
restrictive map $M_A \Gamma_T \to R_A$ describes the restriction
of  arbitrary $\cal T$ state to $\cal A$ degrees of freedom,  i.e. $R_A$ is
$\cal T$ restricted state \cite {B4,B5}. 
In fact, even omitting self-reference arguments such measurement   description looks quite natural. Really, the information acquired by $\cal A$ during $\cal S$ measurement related to its internal state, i.e. the state defined on $\cal A$  degrees of freedom.
%

It was shown that due to such restrictions, if for two arbitrary $\cal T$
states $\Gamma_T,\Gamma'_T$ their restricted  states $R_A, R'_A$
coincide, then for $\cal A$ such $\cal T$ states are
indistinguishable, for any nontrivial $\cal T$
at least one pair of such $\cal T$ states exists \cite {B4,B5}. 
  For classical systems $R_A$ can be calculated directly and  incompleteness of $\cal T$
  state description by $R_A$ has the
  obvious  origin: since $\cal A$ is only the part of $\cal T$,
 the number of $\cal A$ degrees of freedom is always less  than this number for $\cal T$,
 hence $\cal A$ can't discriminate all possible $\cal T$ states $\cite {B4}$. In quantum case, standard QM formalism doesn't permit to define $M_A$ map unambiguously,  it leaves significant freedom for such map  choice. As plausible 'ad hoc'  ansatz of such $\cal T$ state restriction to $\cal A$ the partial trace over $\cal T$ residue usually applied  \cite {B4,B5}.  For given $\cal T$ system such residue is $\cal S$,  in other cases it can include also $\cal A$ environment,  etc..  In particular, for Coleman-Hepp model such $\cal T$ restriction for $\cal T$ state of eq. (\ref {Z2}) results in  $\cal A$ state
\begin {equation}
               R_A=w_1|A_1\rangle\langle A_1|+w_2|A_2\rangle\langle A_2|      \label {Z4}
\end {equation}
for $w_{1,2}=|e_{1,2}|^2$. Then, both for  $\cal P$ state superposition and $\cal P$ gemenge their restricted $\cal T$ states would coincide,
so that corresponding  statistical states  with the same $\bar{S}_z^0$ value  are indistinguishable for $\cal A$.  
For incoming $\cal P$ gemenge $|A_{1,2}\rangle$ eigenvalues  are $\cal A$ objective properties \cite {B1},  so that $\cal A$ final state corresponds to stochastic $|A_{1,2}\rangle$
individual state ensemble. However, for $\cal P$, $\cal A$ entangled state of eq. (\ref {Z2}) and $w_{1,2}\ne 0$ such conclusion will be wrong, the reason is that beside given $R_A$ ansatz there many other state  ensembles described by the same
density matrix $R_A$, however, for such ansatz it can be decomposed in many ways into a convex sum of  pure state projectors \cite {Bra,Bel}. It means that many different $\cal A$ individual  state ensembles can be described by the same statistical state;   example is  given below.
Hence  the choice of partial trace over residue as restrictive map to $\cal A$ doesn't lead to  decomposition of resulting statistical $\cal A$ state into unambiguous ensemble of individual states, corresponding to individual event outcomes, and so can't be accepted as consistent description of $\cal T$ measurement.
This is, in fact,  well-known problem of ignorance interpretability of mixed state  often referred as the "problem of preferred basis", which is in the heart of so-called QM witnessing interpretation \cite {B1,Lah}. 

   



\section { Quantum Measurements in Operator Algebra Formalism }

Here   measurement process will be  considered in operator algebra framework
 \cite {Em4,Bra}.   Algebraic QM  now universally acknowledged
as consistent generalization of standard (Dirac) QM formalism, it was applied successfully to many quantum problems \cite {Bra,Str}. In particular, such formalism exploited for  collective system description, for which standard  QM formalism doesn't work effectively. The examples are spontaneous symmetry breaking, phase transitions, superconductivity or Heisenberg ferromagnetic. In algebraic formalism the basic structures are the system observable algebra $\cal U$, corresponding C$^*$-algebra  and system Hamiltonian, together they  define the system state space $\cal X$ which in some cases can differ from standard QM Hilbert space \cite{Em4,Str}. Typical $\cal X$ example is tensor product of many Hilbert spaces
 $\cal X$$=\cal H$$_1*$$ \cal H$$_2*...$, such state structure appears for mentioned  systems.
 Resulting algebraic states $\{\varphi_i\}$ are functionals on $\cal U$ observables, they correspond to QM statistical states \cite {Em4,Bra}.
If $\varphi_i$ are pure states, then
 Dirac bra, ket notations will be  used here for them.



Concerning the measurement problem,   Emch and coauthors have shown that for  the measurement  of arbitrary system $\cal S$ its state,   characterizing measurement outcome, in general can differ from the state defined on its complete observable algebra $\cal {U}$$^S$ \cite {Em4,Em5}.
 Namely, if some $\cal S$ observables,  due to principal or technical limitations, are unavailable for given measurement set, then $\cal S$ state can be described consistently as the partial state $\varphi^m$ defined on corresponding $\cal U$$^S$ vectorial subspace $\cal M$,  which includes
  unit $I$ and available $\cal S$ observables, so that  $\cal M$ is $\cal U$$^S$ restriction to this set; some $\cal M$ can be $\cal U$$^S$ subalgebras  \cite {Em4,Dix}.
 An arbitrary  state $\varphi^m$ defined on such  $\cal M$,   can be extended properly to some state $\varphi$ defined on
 complete $\cal S$ algebra $\cal U$$^S$, yet such extension, in principle, can be nonunique \cite {Em5,Dix}. In addition, for any such $\varphi^m$ it should exist $\varphi$ on $\cal U$$^S$ which restriction to $\cal M$
coincides with $\varphi^m$, the same relations hold for pure $\cal S$ states defined on $\cal M$ and $\cal U$$^S$  \cite {Em4,Seg}.   
However,  in Emch formalism
$\cal S$ measurement process was treated in standard QM measurement axiomatic, i.e. resulting in its probabilistic  outcomes for available  $\cal S$ observables according to Born postulate \cite {Em4,Em5}. Here it will be  modified to incorporate   self-referential approach to measurements formulated in sect. 4. According to it, 
it will  be considered  for  the  measuring system $\cal T$, which consists of studied system $\cal S$ and  measuring apparatus $\cal A$. As was argued, in the self-reference framework  the restricted $\cal T$ states, which describe the measurement outcome,  defined by $\cal T$ state  mapping to $\cal A$  degrees of freedom \cite {B5}. 
Meanwhile, in quantum case  $\cal A$  degrees  of freedom correspond to its observables and so defined by its observable algebra $\cal U$$^A$ \cite {Bra,Str}.   Hence, it's reasonable to admit that in algebraic QM the corresponding $\cal A$ states, i.e. restricted $\cal T$ states  stipulated by restriction of $\cal T$  observable algebra $\cal U$$^T$ to its $\cal A$  subalgebra $\cal U$$^A$. Therefore, they 
 can be described as the partial $\cal T$ states  on $\cal U$$^A$.  
In particular, for Coleman-Hepp  model  $\cal T$  partial states  defined on given $\cal A$ algebra $\cal {U}$$^A$  corresponding to $\cal A$ spin chain model described in sect. 2.
 It's instructive to study also $\cal U$$^T$ restrictions to all feasible  $\cal {U}$$^A$ subspaces  $\{\cal {M}$$_j\}$,
  because it's possible that some $\cal A$ observables aren't involved in the measurement description  for technical or physical reasons, and so in place of 
   $\cal {U}$$^A$ the proper partial $\cal T$ states would be defined on some  its vectorial subspace.
   
For Coleman-Hepp model  the  observable algebra $\cal U$$^T$ of measuring system $\cal T$  includes $\cal P$
spin subalgebra  $\cal U$$^P$, apparatus $\cal A$ subalgebra $\cal {U}$$^A$ and vectorial subspace
$\cal M$$^{P,A}$ which containes $\cal P$, $\cal A$  joint observables \cite {B1,Es2}.   
For the start  we'll consider this model for $N=1$, so that $\cal A$ includes just one element $\cal C$$^1$ and   $|A_{1,2}\rangle=|s^1_{1,2}\rangle$. In its framework $\cal U$$^P$ is
spin-$\frac{1}{2}$ algebra generated by unit $I$ and three
conjugated observables $\sigma^0_{x,y,z}$ \cite {Em4}. So $\cal
U$$^P\equiv\{G=g_0I+g_1\sigma^0_x+g_2\sigma^0_y+g_3\sigma^0_z\}$
with arbitrary real $g_{0,...,4}$ parameters, it defines $\cal P$ 
state space which coincides with Hilbert space $\cal H$$_s$ of
spin-$\frac{1}{2}$ particle. In the same
vein,  $\cal U$$^A$ subalgebra  for $N=1$ has analogous
structure and
  generated by $I$ and three conjugated observables
$ \sigma^1_{x,y,z}$. It defines the state space $\cal H$$_A$
which is equivalent to  $\cal H$$_A$ of sect. 2 for $N=1$. Plainly, $\cal T$ state evolution is the same in both formalisms
resulting in $\cal T$ final state of eq. (\ref {Z2}). Its 
partial states will be calculated  here for   $\cal U$$^A$  and  its vectorial subspaces, in particular,
  the  partial $\cal T$ state on $\cal U$$^A$ after $\cal P$ measurement
 \begin {equation}
\varphi^A=w_1\varphi_1+w_2\varphi_2  \label {Z11}
\end {equation}
where $\varphi_{1,2}=|A_{1,2}\rangle\langle A_{1,2}|$. 
Hence the  partial state $\varphi^A$
defined on $\cal {U}$$^A$ subalgebra is equivalent to restricted
 state $R_A$ of eq. (\ref {Z4}) with the same $w_{1,2}$,
 analogously to it, $\varphi^A$ also admits nonunique decomposition into  $\cal A$ pure states, its example given below.  By the slight abuse of notations  below $\varphi^A$
 called $\cal A$ partial state, despite that   it's, in fact, $\cal T$ partial state.  

 For this $\cal A$ model   $\cal U$$^A$ vectorial subspaces, involved in measurement process, are  generated by $I$ and some   $\cal A$ observables.
 All of them should include $\sigma^1_z$ observable as generator, because, as follows from eq. (\ref{Z2}), its expectation value defines the  measurement outcome, in particular, $\cal P$ eigensate individual  measurements result in $S^A_z$ eigenvalues which are objective values \cite {B1}.  We'll proceed with their study, considering consequently all such subspaces $ \cal M$$_j \subseteq \cal {U}$$^A$ and
 enlarging  their generator number  one by one.
 Let's start from the  subspace $\cal M$$_1$ that contains just one observable; our formalism premises demand that its expectation value should describe $\cal P$ measurement outcome, however,  there is only one vectorial subspace
$\cal M$ which satisfies to formulated conditions,  it's one-dimensional    
$\cal U$$^A$ subalgebra $\cal U$$^m$.
Its structure is
$\cal U$$^m\equiv \{G=g_0I +g_1\sigma^{ 1}_z\}$ with arbitrary real $g_{0,1}$, 
in this case,
$\sigma^1_z$ eigenstates $|A_{1,2}\rangle$, which correspond to $\cal A$ pointer basis,  exhaust its pure state spectra. Therefore,
now $\varphi^A$ decomposition into pure states can include only $|A_{1,2}\rangle\langle A_{1,2}|$ components and so  
is unique. It corresponds to classical probabilistic ensemble with
$s^1_z=\pm \frac{1}{2}$ outcomes  with probabilities $w_{1,2}$ \cite {B1,Bel}.

In case of two observables as generators one should consider  the set of linear subspaces $\cal M$$_2(Q)$ generated by $I, \sigma^1_z$ and some conjugated  observable $Q$. They would define the corresponding pure state set $\cal L$$(Q)$. For $\cal M$$_2(Q)$ its arbitrary observable $Q^m$  can be described as $Q^m=\lambda_1\sigma^1_z+\lambda_2 Q$ with  arbitrary real $\lambda_{1,2}$. Correspondingly, all $\cal L$$(Q)$ elements are  eigenstates of some $Q^m$ observable.  
  For given $\cal U$$^A$ such $Q$ can be chosen as
\begin {equation}
 Q(\alpha)=\cos \alpha\sigma_x^1+\sin \alpha\sigma_y^1  \label {Z77}
\end {equation}
  for  $0\leq \alpha\leq 2\pi$, so that   each $\cal M$$_2(Q)$ characterized by  fixed parameter $\alpha$. 
 For such vectorial subspace   its corresponding pure state set $\cal L$$(Q)$, which is $\cal H$$_A$ subset, permits  to perform $\varphi^A$ nonunique decomposition \cite {Hug,Bel}. To demonstrate it, let's take   two   positive variables $r_1, r_2$ such that
$$
     \frac{r_1^2}{r_2^2} < \frac{w_1}{w_2} \quad \mbox{and} \quad r_1^2+r_2^2=1
$$
    for $w_{1,2}$ of eq. (\ref {Z11}) so that $r_{1,2}$ can variate within these limits.
 For given $\cal M$$_{2}(Q)$ and corresponding $\alpha$, let's choose the following set of pure states and weights:
$$
     |\psi^{\alpha}_1\rangle=|A_1\rangle, \qquad  w'_1=w_1-\frac{r_1^2}{r_2^2}w_2
$$
\begin {equation}
 |\psi^{\alpha}_2\rangle= r_1|A_1\rangle+r_2 e^{i\alpha}|A_2\rangle, \quad w'_2=\frac{1}{2r_2^2}w_2 \label {Z9}
\end {equation}
$$
   |\psi^{\alpha}_3\rangle= r_1|A_1\rangle-r_2 e^{i\alpha}|A_2\rangle, \quad w'_3=w'_2
$$
As follows from simple algebra 
any $|\psi^{\alpha}_{1,...,3}\rangle$ is eigenstate of some $\cal M$$_2(Q)$ observable $Q^m$ and so
belongs to $\cal L$$(Q)$; hence it's also superposition of some $\sigma_z^1$, $Q(\alpha)$ eigenstates.
For these partial states
\begin {equation}
               \varphi^m (\alpha) = \sum_{i=1}^{3} w'_i |\psi^{\alpha}_i\rangle\langle\psi^{\alpha}_i|
                                        \label {Z6}
\end {equation}  
For arbitrary $r_{1,2}$ within described limits  one obtains that $\varphi^m(\alpha)=\varphi^A$
 of eq. (\ref{Z11}), i.e.  $\varphi^A$ decomposition isn't unique for such subspace $\cal M$$_2(Q)$.
 In fact, any such $\cal M$$_2(Q)$ is real algebra of observables with symmetric $AB+BA$ product; $\cal L$$(Q)$ is real Hilbert space \cite {Em4}. Next, consider the states defined on 
 complete $\cal U$$^A$ algebra, corresponding example of  $\varphi^A$ nonunique decomposition on $\cal H$$_A$  described also by eq. (\ref {Z9})  but with  $\alpha$ varied from $0$ to $2\pi$  \cite {Bel}. Plainly, such decomposition is applicable also for restricted $R_A$ state of eq. ({\ref{Z4}).  Hence there is just one 
 $\cal U$$^A$ vectorial subspace with suitable properties which induces unique $\varphi^A$ decomposition, this is its one-dimensional subalgebra $\cal U$$^m$.   

In fact, due to $\cal P$, $\cal A$ entanglement,  the complete $\cal A$ state can be defined on $\cal T$ algebra $\cal U$$^T$ only, whereas   $\cal U$$^A$, $\cal M$$_2(Q)$ and  $\cal U$$^m$ define just different $\cal T$ partial states.   
Hence both $\cal U$$^A$ and all considered $\cal A$ vectorial subspaces formally can be considered  at equal ground, so that, in principle, partial states defined on  each of them can describe $\cal A$ measurement outcome for $\cal P$ ensemble.
However, any measurement process consists of individual events and so  the proper
$\cal A$ partial state should present their unambiguous  description and provide correct $S^A_z$ expectation value.    
Meanwhile,  of all such $\cal A$ subspaces there is just one subspace -  one-dimensional subalgebra $\cal U$$^m$,
which provides unambiguous decomposition of $\cal A$ statistical states into pure states and thus
corresponds to consistent  ensemble of individual states. For
the rest of  them, including $\cal U$$^A$, $\cal A$ state decomposition into individual state ensemble is nonunique, therefore
  $\cal A$  state decompositions  defined on such subspaces  can be qualified as unphysical. In other words, statistical $\cal A$ state described properly both on $\cal U$$^A$  algebra, as well on arbitrary $\cal M$$_2(Q)$, but such state doesn't permit the proper measurement description on event by event basis. The complete and consistent final state description in terms of $\cal A$ individual states is possible only for states defined on $\cal U$$^m$. 
   Note that analogous situations are well-known in QFT, some QFT problems have multiple formal
  solutions, of them the proper one selected from physical consistency reasons \cite {Em4,Str}.


As follows from eq. (\ref{Z2}), in this model if $\cal A$ measures $\cal P$ up, down eigenstate $|s^0_{1,2}\rangle$, then its  state  will be $S_z^{ A}$  eigenstate $|A_{1,2}\rangle$ with eigenvalues $s^1_{1,2}= \pm \frac{1}{2}$ correspondingly. These $S_z^{ A}$  eigenvalues are  real $\cal A$ properties \cite {B1}, so it's reasonable to admit that $\cal A$ percepts these eigenstate difference as
difference of two natural numbers or yes/no propositions. In distinction,
 by itself  arbitrary $\varphi^A$ of eq. (\ref{Z11}) can't be interpreted as $\cal A$ individual state. Really, $|A_{1,2}\rangle\langle A_{1,2}|$ are individual $\cal A$ states, their distinction stipulated by their $S^A_z$ eigenvalues $s^1_{1,2}$, which are objective and can be discriminated in single event. Yet for any pair $\varphi^A$, $|A_{1,2}\rangle$ for $w_{1,2}\ne 0$ no $\cal A$  observable
exists, which would permit to perform such  discrimination by their eigenvalues. Hence the only reasonable conclusion is that correct $S^A_z$ expectation value for such pure ensembles can be realized only via $|A_{1,2}\rangle$ fundamentally
stochastic measurement outcomes.  
 It's notable that for arbitrary  incoming $\cal P$ states only $S^A_z$ final expectation values  can variate 
for that measurement model, but  $S^A_{x,y}$   expectation  values $ \bar{S}^A_{x,y}=0$.  Hence,  considered measururement outcome description involves effectively only one-dimensional $\cal A$ subalgebra $\cal U$$^m$,
application of $\cal U$$^A$ or its other vectorial subspaces seems excessive.

.

\section {Microscopic Measurement  Models in Algebraic Formalism}

Here algebraic formalism will be applied to more realistic measurement models, which attempt to describe macroscopic devices functioning. First, simple variant of  decoherence model, proposed  by Zurek, will
be considered \cite {B1,Es2}. Such model also deals with measurement of particle $\cal P$ spin-$\frac{1}{2}$  projection by apparatus $\cal A$, but takes into account $\cal A$ interaction with its environment $E$.  In this model  the  measurement process consists of two stages; first,
 $\cal P$ interacts with measuring 
 apparatus $\cal A$, when this interaction seized, $\cal A$ starts to interact with  $E$ which consists of $L$ independent two-level systems (atoms), the $k$th of them has a two-dimensional Hilbert state space $\cal H$$_k$ with basis $|u^k\rangle, |d^k\rangle$ \cite {Z10}. Its initial normalized state in this basis is $\{\alpha^k_1, \alpha^k_2\}$ with complex $\alpha^k_{1,2}$ chosen at random.
  $\cal A$  also has two-dimensional Hilbert space with $|A_{1,2}\rangle$ basis
and so is equivalent to one of Coleman-Hepp model for $N=1$. Thus, in such model $\cal P$, $\cal A$ and $E$ are parts of measuring system $\cal T$ with corresponding observable algebra $\cal U$$^T_E$.
   Initially , the measured particle $\cal P$
is in pure spin state of eq. (\ref{AA99}), when $\cal P$, $\cal A$ interaction seized, in Zurek model it
    results in the entangled $\cal P$, $\cal A$  state
   \begin {equation}
   \psi(s^0,s^{1}) =e_1|s^0_1\rangle|A_1\rangle+
    e_2|s^0_2\rangle|A_2\rangle   \label {Z33}
   \end {equation}
   analogous to $\cal P$, $\cal A$    state of eq. (\ref{Z2}). Then, after $\cal A$, $E$ interaction finished,  $\cal T$ system state becomes
$$
\Phi(t) =e_1|s^0_1\rangle|A_1\rangle \prod_{k=1}^L \otimes[\alpha^k_1 e^{ig_kt}|u^k\rangle  +\alpha^k_2 e^{-ig_kt}|d^k\rangle]+
$$
\begin {equation}
  +e_2|s^0_2\rangle|A_2\rangle \prod_{k=1}^L \otimes [\alpha^k_1 e^{-ig_k t}|u^k\rangle+ \alpha^k_2 e^{ig_k t}|d^k\rangle ]  \label {Z22}
   \end {equation} 
where coupling constant $g_k$ of $k$th atom with $\cal A$ is chosen at random \cite {Z10}. The density matrix $\rho_r$ of $\cal P$, $\cal A$ system is then obtained by    taking the trace over $E$,  for $L\to \infty$ in short time $\rho_r$ nondiagonal elements become quite small \cite{Z10}. Decoherence theoretists claim that it means that apparatus pointer is in definite but unknown to observer position corresponding to one of $S^A_z$ eigenvalues. 
Yet it was demonstrated that this conclusion is incorrect, in fact, $\cal P$, $\cal A$ system still is in more complicated, but pure state of eq. (\ref{Z22}), entangled with $E$ elements, so can't correspond to objective $S^A_z$ values \cite {B1,Es2}.

Meanwhile,  in algebraic QM formalism  $\cal A$ partial  states for Zurek model defined by $\cal U$$^T_E$ restriction to its subalgebra $\cal U$$^A$, as the result, they   are equivalent to  partial $\cal A$ states with the same $e_{1,2} $ for Coleman-Hepp model  with $N=1$ considered in sect. 5, because their $\cal A$ subalgebras coincide. Really,  
 its partial $\cal A$ state  is the functional defined on
$\cal A$ algebra $\cal U$$^A$ and it coincides with $\varphi^A$ of  eq. (\ref{Z11}). As was shown, such state  possesses nonunique decomposition into pure states on $\cal A$ algebra $\cal U$$^A$. In the same vein,  its partial states on arbitrary $\cal A$ vectorial subspace  can be calculated analogously to ansatz of  sect. 5, and it would give the same result, i.e. there is just one  unique partial state decomposition induced by 
one-dimensional $\cal A$ subalgebra $\cal U$$^m$. It corresponds to probabilistic mixture of pointer observable eigenstates;  the partial $\cal A$ states defined on other $\cal A$ subspaces, including $\cal U$$^A$, admit nonunique decomposition. Hence, alike in Coleman-Hepp model for $N=1$, the  only consistent solution is $\cal A$ partial state $\varphi^A$ defined on subalgebra $\cal U$$^m$, which corresponds to $|A_{1,2}\rangle$ probabilistic ensemble of individual states.


 For  Coleman-Hepp model  with  $N\ge 2$ it's instructive to start from the case $N=2$, because the main results for it will be applicable also for arbitrary $N$ and  generator number $l$. Then, $\cal A$ spin chain includes  two elements $\cal C$$^{1,2}$, so that for incoming $\cal P$ state  of eq. (\ref{AA99}) $\cal A$ final partial state
\begin {equation}
\varphi_2^A =  w_1| A_1\rangle \langle A_1|+
w_2|A_2\rangle \langle A_2|= w_1| s^1_1\rangle \langle s^1_1||s^2_1\rangle\langle s_1^2|+
w_2| s^1_2\rangle \langle
s^1_2||s^2_2\rangle \langle s^2_2|
\label {Y1}
\end {equation}
Let's study which class of $ \cal A$ observables and  corresponding vectorial subspaces induces $\varphi^A_2$ nonunique decompositions.
 Plainly, $\cal C$$^1$ 
or $\cal C$$^2$ individual  observables are of no interest for that, because their eigenstate
projectors can't be part of $\varphi_2^A$ decomposition. Hence only joint 
$\cal C$$^{1,2}$  observables should be studied as possible $\cal U$$^A$ vectorial subspace generators.
For $\cal A$ Hilbert space $\cal H$$_A$
the convenient basis for joint $\cal C$$^{1,2}$ observable eigenstate description can be chosen as 
$$
|\eta_{1,2}\rangle =|s^1_{1,2}\rangle |s^2_{1,2}\rangle ; \,\,
|\eta_{3,4}\rangle= |s^1_{1,2}\rangle |s^2_{2,1}\rangle\ 
$$
 i.e. $|\eta_{1,2}\rangle=|A_{1,2}\rangle$. Any eigenstate of arbitrary $\cal C$$^{1,2}$ joint observable can be written in this basis as
$$
    |\psi^{_\eta} \rangle =\sum_{i=1}^4 g_i|\eta_i\rangle
$$
for  complex $g_i$ with $\sum |g_i|^2=1$.
 Plainly, any proper vectorial subspace should include  $S^A_z$ observable of eq. (\ref{Z55}), which expectation value describes $\cal P$ measurement outcome.  Hence  such one-dimensional subspace  can be only subalgebra
 $\cal U$$^m$ generated by $I, S^A_z$ observables. For $N=1$ such $\cal U$$^m$ was studied in sect. 5., analogously to it, for $N=2$ it induces unique  $\varphi_2^A$ decomposition, which corresponds to $|A_{1,2}\rangle$ probabilistic ensemble of individual states. Really, for such subalgebra only $S^A_z$ eigenstate projectors $|\eta_{1,2}\rangle \langle\eta_{1,2}|$ provide correct $\varphi_2^A$ decomposition.
Now let's consider vectorial subspace $\cal M$$_2(K)$ generated by $I, S^A_z$ and an arbitrary joint $\cal C$$^{1,2}$   observable $K$. They define the corresponding pure state set $\cal L$$(K)$.  For $\cal M$$_2(K)$ its arbitrary observable $Q^k$  can be described as $Q^k=\lambda_1 S^A_z+\lambda_2 K$ with  arbitrary real $\lambda_{1,2}$. Correspondingly, $\cal L$$(K)$ elements are  eigenstates of some $Q^k$ observable. 

 As follows from simple algebra, analogous to eq. (\ref{Z9}), for any $K$ eigenstate  $|\psi^{k}\rangle$,
  which decomposition  in that basis includes $|\eta_{3,4}\rangle$, its projector would contain one or two $|\eta_{3,4}\rangle \langle\eta_{3,4}|$ terms with positive weights $w_{3.4}$, yet it's incompatible with $\varphi_2^A$ ansatz  of eq. (\ref{Y1}). Such observables called below passive.  
   For example, for $\cal M$$_2(S^A_{x,y})$  such $|\eta_{3,4}\rangle \langle\eta_{3,4}|$ terms would be provided by $S^A_{x,y}$ eigenstates. Therefore, such $\cal M$$_2(K)$ also induces unique $\varphi^A_2$ decomposition, but it's equivalent to one induced by $\cal U$$^m$, being just  $w_{1,2}|\eta_{1,2}\rangle \langle\eta_{1,2}|$ sum provided by $S^A_z$ eigenstates.  Hence for such $\cal M$$_2(K)$ its subalgebra $\cal U$$^m$ is minimal $\cal A$ vectorial subspace which induces unique $\varphi^A_2$ decomposition.  If arbitrary $\cal M$$_l$,
   beside $I$ and $S^A_z$ includes also $l-1$ passive observables, it also would induce the same unique $\varphi^A_2$ decomposition.


It prompts  that for arbitrary observable $K$, for which $\cal M$$_2(K)$ induces nonunique $\varphi_2^A$ decomposition, $K$ eigenstates should be some $|A_{1,2}\rangle$ superpositions, because they are orthogonal to $|\eta_{3,4}\rangle$. 
Yet analogous states  were considered   in eq. (\ref{Z24})  for $\cal P$, $\cal A$  system with $N=1$, for $N=2$
 such $\cal  A$ states 
 \begin {equation}
|\psi^{u}\rangle=u_1|A_1\rangle+u_2|A_2\rangle 
\end {equation}   
for arbitrary complex $u_{1,2}\neq 0$ and $|u_1|^2+|u_2|^2=1$. It means that corresponding  $K$
observables are similar to $G_1^s(c_{1,2})$ observables of eq. (\ref{Z88}) changing indexes from $0,1$ to $1,2$  correspondingly.
For arbitrary $N\ge 2$ they are
 \begin {equation}
G^c_N(c_{1,2})=c_1\prod_{i=1} ^N |s^i_{1}\rangle \langle s^i_{2}|+ c_2\prod_{i=1} ^N|s^i_{2}\rangle \langle s^i_{1}|
\label {Z89}
\end {equation}
 for complex $c_{1,2}$ with $c_1 =c_2^*$, $|c_{1,2}|=\frac{1}{2}$. 
In fact, they are interference observables for $\cal C$$^{1,...,N}$ joint  states, their class denoted $\{ G^c_N\}$.
 Then, for $N=2$ any $\cal M$$_2(Q)$ subspace,
which generators are $S^A_z$ and some  $Q \in \{G^c_2\}$   would induce nonunique $\varphi_2^A$ decompositions described by eq. (\ref{Z9}).
These results demonstrate the existence of multiple nonunique $\varphi_2^A$ decompositions on $\cal U$$^A$.
 Plainly, extension of such  $\cal M$$_2(Q)$  to some $\cal M$$_3$ by inclusion as its generator of some additional $\cal U$$^A$ observable $V$, which doesn't belong to  $\{G^c_2\}$ class, 
wouldn't change the situation, because its eigenstates would contain $|\eta_{3.4}\rangle$ components, and so their projectors can't participate in $\varphi^A_2$ decomposition.  
  Therefore, for $N=2$ both for $\cal A$ algebra $\cal U$$^A$ and  its arbitrary vectorial subspaces only  partial $\cal A$ states defined on its one-dimensional  subalgebra $\cal U$$^m$   admit nontrivial unique decomposition, which describe measurement outcomes for $\cal A$  individual states consistently.


For  $N \ge 3$ situation is analogous,  the partial $\cal A$ final state on $\cal U$$^A$
\begin {equation}
   \varphi_N^A = w_1\varphi^c_1+   w_2\varphi^c_2   \label {Y2}
\end {equation}
where $\varphi^c_{1,2}=|A_{1,2}\rangle\langle A_{1,2}|$ with $| A_{1,2}\rangle$ of eq. (\ref{Z0})
also admits nonunique decompositions on $\cal U$$^A$. $S^A_z$ has $2^N$ eigenstates which constitute complete basis for $\cal C$$^{1,...,N}$ joint states on $\cal U$$^A$. 
The same reasoning as for $N=2$ demonstrates  that
   only the states defined on $\cal U$$^A$ one-dimensional subalgebra
$\cal U$$^m$ generated by $I, S^{ A}_z$ observables, constitute its  nontrivial unique decomposition. It results in consistent measurement outcome description for individual states as probabilistic ensemble of  $| A_{1,2}\rangle$ states. For arbitrary $N$ and any  $\cal U$$^A$ vectorial subspace $\cal M$$_l$, if one of its generators is $S^A_z$, and  $\cal M$$_l$ induces nonunique $\varphi^A_N$ decomposition, then at least one of its generators $Q\in \{G^c_N\}$  described by eq. (\ref{Z89}).  Otherwise, if  all its generators, except $S^A_z$, are passive, then it  induces  $\varphi_N^A$ unique decomposition,
but it's equivalent to decomposition induced by subalgebra $\cal U$$^m$.
 These results demonstrate  the existence of multiple nonunique $\varphi_N^A$ decompositions on $\cal U$$^A$ for arbitrary $N$.

   


\section {Conclusion}


Here algebraic QM formalism  for Zurek  and Coleman-Hepp measurement models was studied, however, one can expect that for  more complicated and realistic  systems with different observable algebra the results for PV observable measurements
can be analogous. In standard  measurement  framework the system Hamiltonian  maps the measured PV observable $V$ to apparatus $\cal A$ (pointer) observable $U$, i.e. for arbitrary measured state after the measurement finished it would give $\bar{U}\sim\bar{V}$ [1-4]. In algebraic QM formalism it will define one-dimensional $\cal A$ subalgebra $\cal U$$^m$ with one generator $ U$, which induces the unique decomposition  of partial $\cal A$ state corresponding to classical statistical ensemble. Plainly, independently of $V$ eigenvalue number such subalgebra would induce the state space, which  performs unique
$\cal A$ partial state decomposition with the same properties corresponding to  probabilistic ensemble of individual $\cal A$ states. As was shown, it stipulated by the impossibility to transfer to $\cal A$ the information about  measured ensemble purity rate simultaneously with information about $V$ expectation value. Here only two kinds of measured state ensembles were considered: the orthogonal binary state superposition and corresponding gemenge. However,
our study  doesn't reveal any complications for applicability of this formalism to measurements of arbitrary quantum ensembles.

It's notable that  by itself Breuer description  of measurement from inside  leaves significant freedom for the  choice
of restricted state ansatz, so that to large extent such choice can be stipulated by its consistency with physical reasons discussed here. In fact, operator algebra seems to be quite natural framework for this approach, because its restrictive map ansatz relates the measuring system and apparatus degrees of freedom, both of them just correspond to their observable algebras. Such 
correspondence results in fundamental stochasticity  of measurement outcomes.
Enlarging the number of elements $N$ in Coleman-Hepp or Zurek model to infinity
doesn't change these results.
Obtained results don't contradict to Von Neuman uniqueness theorem \cite {Bra,Str}, because it
deals with the representations of the same system, but not with  relation between representations of  system and its subsystem exploited here. In general, obtained results support the hypotheses of principal incompleteness of quantum measurement  \cite {M6}.



Described formalism permits to reconsider notorious Wigner's  friend paradox \cite{W9,Fr}. Namely, suppose that apparatus $W$ performs  measurement of binary state superposition, analogous to  $\cal P$ state of eq. (\ref{AA99}).  Born postulate prompts that $W$ should detect it as one of two possible eigenstate, i.e. to be in mixed state.  Apparatus $F$  is in stand-by position, permitting potentially to measure arbitrary $\cal P$, $W$ observables.  From $F$ point of view after the measurement  $W$ stays in the state of superposition of these outcomes. In algebraic QM framework, these two pictures don't contradict to each other, because $W$ and $F$ deal with the states defined on different observable algebras, hence $W$ state for measurement from inside by $W$, can differ for $W$ state for $F$, which can perform $\cal P$, $W$ system  measurement from outside. Note that account of $W$ decoherence by its environment makes
this paradox quite abstract entity, because corresponding interference observables, which can reveal pure/mixed $\cal P$, $W$ ensemble difference
become practically unaccessible  for $F$.  Some novel aspects of self-reference in quantum measurements discussed in \cite {M6,Fr}.

\end {document}